\documentclass[aps,prl,preprint,superscriptaddress]{revtex4-1}
\usepackage{graphicx}
\usepackage{color}
\usepackage{xcolor}
\usepackage{lineno}
\usepackage{amssymb}
\usepackage{setspace}
\renewcommand{\figurename}{{\bf Fig.}}

\usepackage[normalem]{ulem}

\begin{document}


\title{Topology of vibrational modes predict 
plastic events in glasses}

\author{Zhen Wei Wu}
\email[]{zwwu@bnu.edu.cn}
\affiliation{Institute of Nonequilibrium Systems, School of Systems Science, Beijing Normal University, 100875 Beijing, China}
\affiliation{International Center for Quantum Materials,
School of Physics, Peking University, Beijing 100871, China}
\author{Yixiao Chen}
\affiliation{Yuanpei College, Peking University, 100871 Beijing, China}
\author{Wei-Hua Wang}
\affiliation{Institute of Physics, Chinese Academy of Sciences,
Beijing 100190, China}
\author{Walter Kob}
\email[]{walter.kob@umontpellier.fr}
\affiliation{Laboratoire Charles Coulomb, University of Montpellier
and CNRS, 34095 Montpellier, France}
\author{Limei Xu}
\email[]{limei.xu@pku.edu.cn}
\affiliation{International Center for Quantum Materials,
School of Physics, Peking University, Beijing 100871, China}

\date{\today}

\begin{abstract}
The plastic deformation of crystalline materials can be understood by considering their structural defects such as disclinations and dislocations. Although glasses are also solids, their structure resembles closely the one of a liquid and hence the concept of structural defects becomes ill-defined. As a consequence it is very challenging to rationalize on a microscopic level the mechanical properties of glasses close to the yielding point and to relate plastic events with structural properties. Here we investigate the topological characteristics of the eigenvector field of the vibrational excitations of a two-dimensional glass model, notably the geometric arrangement of the topological defects as a function of vibrational frequency. We find that if the system is subjected to a quasistatic shear, the location of the resulting plastic events correlate strongly with the topological defects that have a negative charge. Our results provide thus a direct link between the structure of glasses prior their deformation and the plastic events during deformation.

\end{abstract}

\maketitle

The structural disorder of glasses allows them to have a multitude of properties that are absent in crystals but can be exploited in many technical applications~\cite{Deb01,Bin11,Var19,Wan19}. The downside of this disorder is that it has hindered us to come up with a reliable microscopic description of many of these features
since at present we lack a solid understanding how the atoms are arranged on the level of the particles~\cite{Bin11}.  Prominent examples of these properties are the specific heat of glasses at low temperatures, the electric conduction properties of glasses~\cite{Phi87,Ram22}, or the plastic deformation of the solid upon applied load before the yielding point~\cite{Dyr06}.
While in crystals dislocations and disclinations allow to rationalize the plastic deformation~\cite{kit96}, the structural disorder of glasses makes it impossible to come up with a reasonable definition of a defect and as a consequence it becomes very challenging to establish a connection between the local structure and the yielding of the sample, despite a multitude of efforts~\cite{Fal98,Pen11,Man11,Din14,Pat16,Yan16,Ric20,Rid22}. Although the local structure of systems composed of particles that have a mesoscopic size (colloids, granular systems,...) can be determined on the level of the particles~\cite{Che11,Yan16,Cao18}, the mechanical properties of such materials are dominated by particle-interactions that are mainly of excluded volume type, i.e., very different from the ionic or covalent interactions found in atomic systems. As a consequence our understanding on how the local structure and interactions affect the mechanical properties of materials like metallic or oxide glasses is far from being satisfactory, despite the long history of research on this topic~\cite{Nic18}.

Early studies on this question showed that for simple model systems an increasing strain leads to a destabilization of the local packing and results in a strain activated, instead or thermally activated, stress relaxation~\cite{Mala98}, a picture that was elaborated by Falk and Langer who introduced the notion of shear transformation zones (STZ)~\cite{Fal98}. Many subsequent studies have related these STZ and the plastic events (PE) to structural quantities or a high local entropy, indicating that this concept is useful~\cite{Pen11,Din14,Yan16,Cao18,Mit22}.

By probing the local yield stress in the glass, Patinet {\it et al}  showed that the zones in which this stress is weak correlate well with the spots at which the globally sheared sample shows a plastic event~\cite{Pat16}. In particular it was found that this purely local measure shows a higher predictive power for identifying sites of plastic activity than the structural properties considered so far, thus demonstrating that PE can be predicted from local information and that therefore it is reasonable the attempt to correlate PE with local structural information.

At low temperatures the motion of the atoms is of vibrational nature~\cite{Ler21} and since PEs are related to a local mechanical instability of the particle configuration it can be expected that the latter are associated with quasi-localized soft modes. Various studies support this view, showing that the location of these soft modes are often, but not always, related to the local arrangement of the particles~\cite{Maz96,Sch04,Mal04,Mal04a,Bri07,Wid08,Xu10,Tang10,Man11,Che11,Din14,Rot14,Gar16,Bon19,Bag21}.

In view of the difficulty to identify the relevant structural features that are responsible for the occurrence of PE, recent studies have used machine learning approaches to identify a connection between these two quantities~\cite{Fon22,Sch17,Cub17,Fan21,Rid22}. 
Although these works have demonstrated that this technique is indeed able to predict to some extent the location of PEs, Richard and collaborators has recently shown that the quality of these predictions is not always superior to the one of simple local structural predictors and hence one concludes that at present we are still lacking insight on the relevant quantity that is able to predict PEs~\cite{Ric20}.

Although soft modes show a good correlation with plastic events, one must recall thatyielding is a cooperative phenomenon. Therefore it is important to study not only the effect of individual modes, but instead to consider the total vibrational field given by the weighted sum over all the modes, since this will determine the displacement field of all the atoms. In the present work we hence focus on the eigenvector field for the different modes, and in particular study its topological properties. (See also Ref.~\cite{Bag21} for a somewhat related approach.) We find that the topological singularities of this field, averaged over the modes, are closely related to the plastic events, showing that the vibrational modes allow direct prediction of the location of plastic transformations if the sample is sheared.

{\it System:} 
The system we study is a two-dimensional binary mixture of Lennard-Jones particles, the interactions of which have been truncated at the minimum.
The liquid was equilibrated at a high temperature and then cooled down, at constant volume, below the kinetic glass transition temperature. Using a conjugate gradient procedure we determined the local minimum of the potential energy and subsequently calculated the eigenvectors and vibrational frequencies $\omega$. More  details on the potential and the simulations are given in the Methods and structural data is given in the Extended Data Figs.~\ref{fig_s1_s_of_q} and \ref{fig_s9_g_of_r}.

\begin{figure}[th]
\centering
\includegraphics[width=1.0\linewidth]{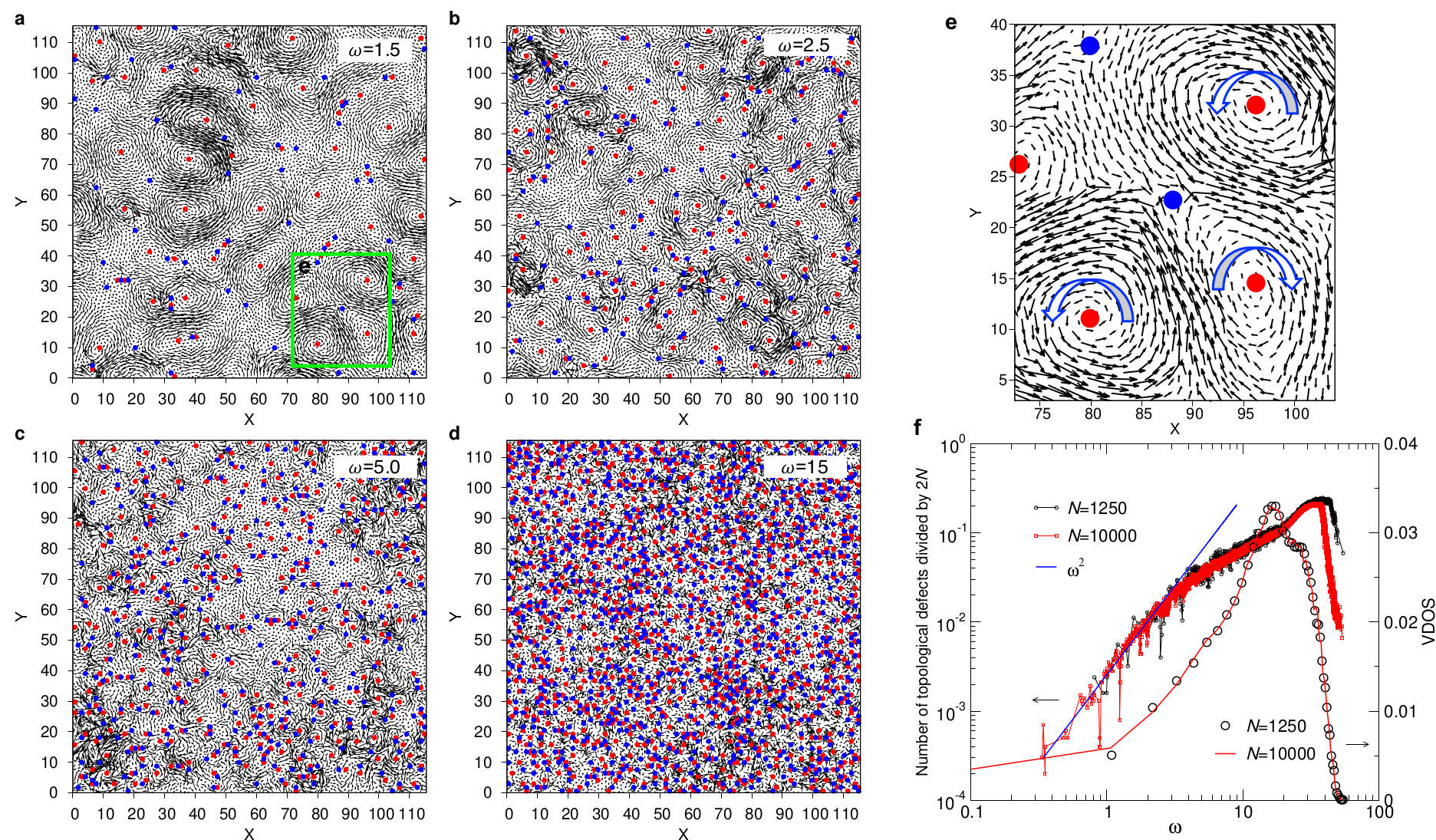}
\caption{
{\bf Normal modes eigenvectors and topological defects.}
{\bf a-d}, Snapshots of the normalized eigenvector at  $\omega$=1.5, 2.5, 5, and 15, respectively.
The green box in panel {\bf (a)} shows the location of the zoom of panel {\bf (e)}.
Note that in {\bf (a)-(d)} the magnitude of the eigen-vectors in different modes
are shown on the same scale (amplified 150 times for visibility).
{\bf e}: Four $+1$ defects (red) and two $-1$ defects (blue). The big arrows mark the chirality of the defects.
{\bf (f)}: The number of topological defects per degree of freedom, 2$N$, as a
function of $\omega$ (left scale) for two system sizes. The solid line is a power-law with exponent 2.0. Right scale: Vibrational density of states.
}
\label{fig1_TD}
\end{figure}

{\it Results:}
In order to comprehend the evolution of the morphology of the eigenvector field with $\omega$, we show in Fig.~\ref{fig1_TD} the vibrational modes for different values of the frequency. (The vibrational density of states is shown in panel {\bf(f)}.) For low frequencies, panel (a), the mode is composed of a multitude of swirls that have roughly the same size. (In Extended Data Fig.~\ref{fig_s3_ev} we show that at very low frequencies the eigenvector does show a regular pattern, as expected for an elastic solid.) By calculating from the eigenvector field the topological charge, i.e., the winding number, we can identify the location of the singularities (topological defects, TD) of the field (see Methods) and these positions are included in the graph as well, with red and blue points marking charges +1 (vortex) and $-1$ (saddle points), respectively; see also panel {\bf (e)}. 

A closer inspection of Fig.~\ref{fig1_TD}{\bf (a)} demonstrates that at low $\omega$ the positions of the TD are not random but correlated: While TDs with the same sign have the tendency to stay away from each other, TDs with opposite sign seem to attract each other and below we will quantify this behavior in more detail. Note that two neighboring swirls with opposite sign of phase chirality (direction of rotation of the vortex) have an interface that is not frustrated, Fig.~\ref{fig1_TD}{\bf (e)}, while if they have the same sign of chirality, they are not compatible with each other and thus at their interface there will be a TD with charge $-$1, Fig.~\ref{fig1_TD}{\bf (e)}.

\begin{figure}[th]
\centering
\includegraphics[width=1.0\linewidth]{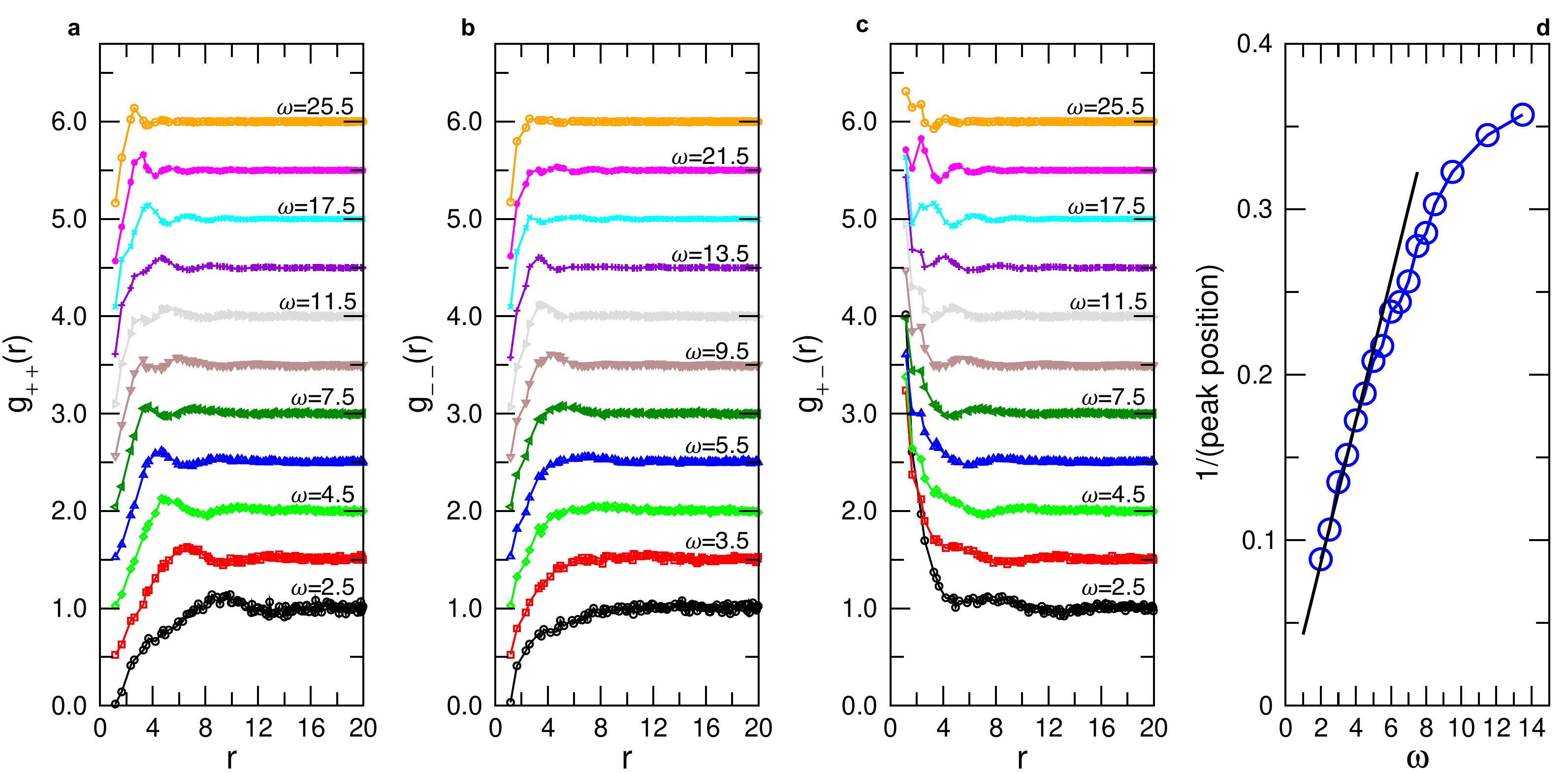}
\caption{
{\bf Structure of the topological defects in the eigenmodes.}
{\bf (a)-(c)}: Pair correlation functions for +/+, $-$/$-$, and +/$-$ defect
pairs for different values of $\omega$. Curves for $\omega>2.5$ are shifted upward by multiples of 0.5. {\bf (d)}: Inverse of the position of the first peak
in $g_{++}(r)$ as a function of frequency. The straight line is a linear fit to the data at low $\omega$.
}
\label{fig2_struTD}
\end{figure}

If $\omega$ is increased, panels {\bf (b)}-{\bf (d)}, it becomes increasingly difficult to identify isolated zones in which the atoms move in an coherent manner, i.e., the swirls become ill-defined. It is, however, still possible to identify the TDs and thus to recognize that their number increases quickly with increasing frequency. Panel~{\bf (f)} shows the total number of TD as a function of frequency and we find that at low $\omega$ this number increases like $\omega^2$. This dependence can be understood by recalling that for a continuous elastic medium the number of maxima/minima of a plane wave with wave-vector ${\bf q}$, i.e., an acoustic mode, increases like $q=|{\bf q}|$. 
If $k$ plane waves having the same $q$ but different orientations are superposed, this will give rise to $k \cdot q$ minima/maxima. Since in $D$ dimensions the number of modes increases like $k \propto q^{D-1}$, the number of minima/maxima scales like $q^D$ and because $\omega \propto q$, this rationalizing the $\omega$-dependence of the number of TD (see also Fig.~\ref{fig2_struTD}(d) and Extended Data Fig.~\ref{fig_s3_ev}).

The positional correlation observed in Fig.~\ref{fig1_TD}{\bf (a)} between the defects can be quantified by means of the corresponding partial radial distribution functions $g_{++}(r)$, $g_{+-}(r)$, and $g_{--}(r)$~\cite{Bin11}, Fig.~\ref{fig2_struTD}. For low $\omega$ one notices that $g_{++}$ has a correlation hole at small $r$, followed by a small peak at $r \approx 9$, signaling the typical distance between these TDs, in agreement with the fields shown in Fig.~\ref{fig1_TD}. With increasing $\omega$ this peak shifts to smaller $r$ and one finds that at small and intermediate $\omega$, its location is proportional to $\omega$, see panel {\bf (d)}, which is consistent with the $q$-dependence of the acoustic modes discussed above. This proportionality holds up to $\omega \approx 4.5$, at which point it crosses over to a weaker $\omega$-dependence. In this range of $\omega$, $g_{++}$ displays several wiggles, indicating that the TD with charge +1 form  a structure that is similar to the one of a liquid, i.e., these TDs have the tendency to repel each other. For $g_{--}(r)$, panel {\bf (b)}, we find no nearest neighbor peak at low $\omega$, showing that these TD behave like an ideal gas. Only if the frequency reaches $\omega\approx 7.5$ one notices a peak at around $r=5$, but at larger $r$ no further wiggles are observed. This indicates that negative TDs are clustered but do not form a structure that is reminiscent to the one of a liquid. If $\omega$ is increased even more, the correlation function becomes again flat (except for a correlation hole at $r=0$), showing that at high frequencies these TDs become completely uncorrelated. These observations for $g_{--}(r)$ indicate that the relative arrangement of the $-1$ defects is significantly less pronounced than the one of the $+1$ TD and below we will discuss the origin of this behavior.

The correlator $g_{+-}(r)$, panel~{\bf (c)}, shows a dependence on $r$ and $\omega$ that is very different from the two other correlators. Independent of $\omega$ one finds a strong peak at small $r$'s, demonstrating that the positive and negative TD attract each other. In view of this strong correlation one can conclude that the positive TD form a liquid-like structure and that it is the attraction between the positive and negative TD which makes that the latter become correlated as well, although to a lesser extent, in agreement with the data in panel~{\bf (b)}.


\begin{figure}[th]
\centering
\includegraphics[width=1.0\linewidth]{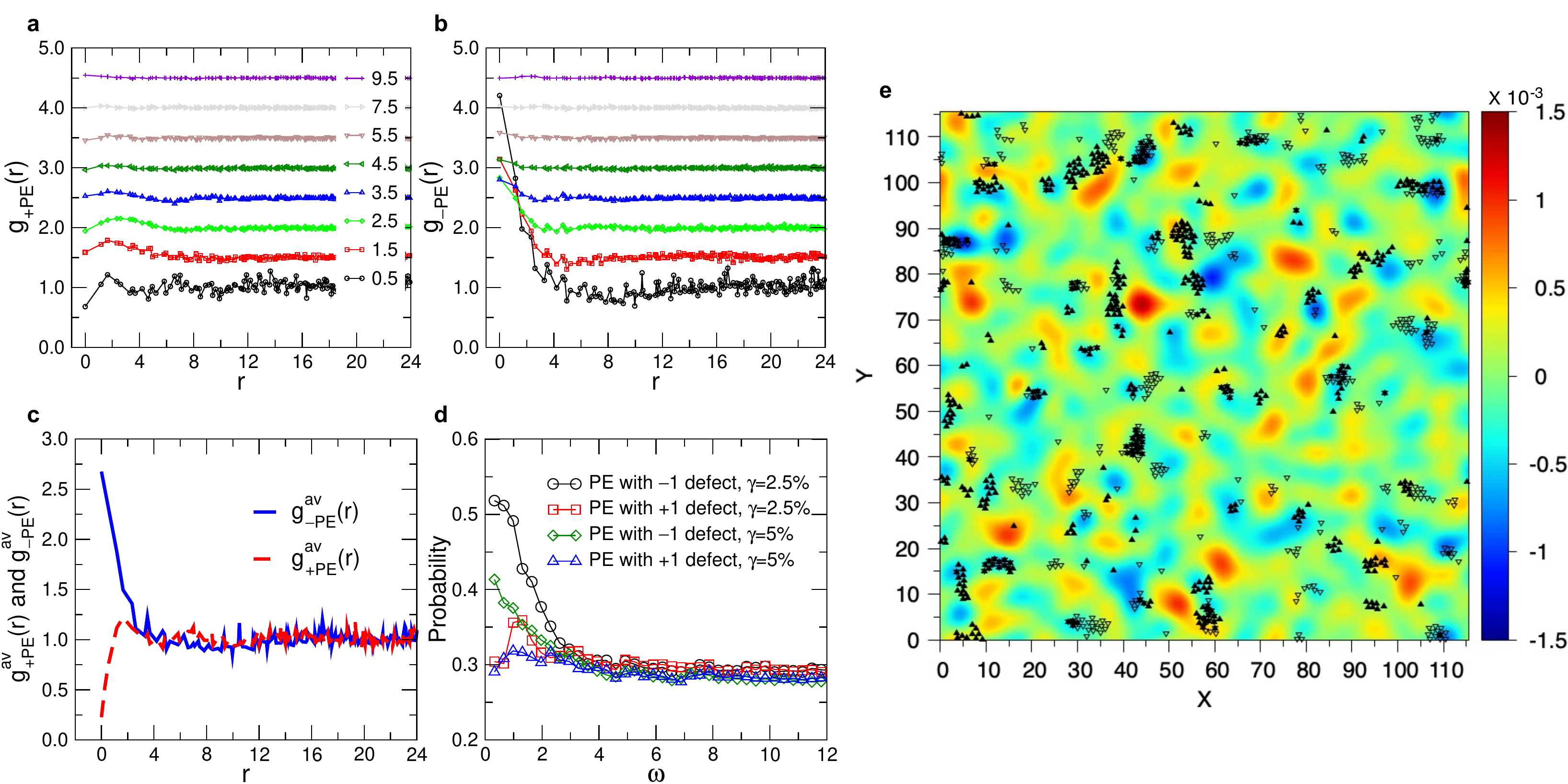}
\caption{
{\bf Spatial correlation between the location of topological defects and the plastic events.}
{\bf (a)} and {\bf (b)}: 
Correlation functions between plastic events at strain $\gamma=2.5$\% and topological defects with positive charge, {\bf (a)}, and negative charge, {\bf (b)}, for different frequencies.
{\bf (c)}: Weighted sums over $\omega$ of $g_{+{\rm PE}}(r)$ and $g_{-{\rm PE}}(r)$. The weights are $\omega^{-2}$ and all modes up to $\omega=3.5$ have been taken into account.
{\bf (d)}: Probability that a PE has a distance less than 1.6 from a negative/positive PE (see legend). Two values of the strain $\gamma$ are shown. 
{\bf (e)}: Charge density field of the TDs (color map) and of the PE (symbols) at strain $\gamma=0.025$. The density field, in units of charge per unit area, has been obtained by averaging the TDs over all frequencies up to $\omega_{\rm max}=3.5$, using a weight factor $\omega^{-2}$, and a subsequent Gaussian smoothing function of width 4. PE resulting from a shear in the positive and negative $x$-direction are marked by open and filled symbols, respectively.
}
\label{fig3_TDPE}
\end{figure}

Since TDs are singularities in the field of the eigenvector, it can be expected that they are directly related with the heterogeneous mechanical response of the system under shear. To explore this connection we have determined the plastic instabilities of the sample if it is put under an athermal quasi-static pure shear, see Methods for details. For a sheared configuration at strain $\gamma$ we calculate $D^2_{\rm min}$, the field of non-affine displacement of the particles between two consecutive configurations  (having a difference in strain of 0.05\%)~\cite{Fal98}. After a drop in the stress, signaling that a plastic event has occurred, we identify the particles that have a high $D^2_{\rm min}$ (top 5\%, see Extended Data Fig.~\ref{PDF_d2}), and associate these particles to a PE.
The location of these PEs can be correlated with the position of the TDs by means of a corresponding radial distribution function $g_{\alpha \rm PE}(r)$, 
where $\alpha \in \{\pm 1\}$. In Fig.~\ref{fig3_TDPE}{\bf(a)}-{\bf (b)} we show the resulting correlation for different values of $\omega$ and one recognizes that $g_{+ \rm PE}(r)$ is basically flat, implying that there is only very little correlation of the PE with the positive TDs, and this holds for all frequencies. The only noticeable feature in the curve is the small peak at around $r=2$ if $\omega$ is small, the origin of which is discussed below.

In contrast to this, $g_{-{\rm PE}}(r)$ shows at small and intermediate values of $\omega$ a marked peak at small $r$, panel {\bf (b)}, demonstrating that PE are closely related to low-frequency TDs with negative charge. This results rationalizes also the presence of the small peak found at $r\approx 2$ in $g_{+{\rm PE}}(r)$ since from Fig.~\ref{fig2_struTD}{\bf (c)} we know that TDs with positive and negative charges are significantly correlated and hence the correlation between negative TDs and PEs will induce a correlation between positive TDs and PEs.

Panels {\bf (a)} and {\bf (b)} show that TDs and PEs are correlated if $\omega$ is less than $\omega_{\rm max}=3.5$, while for higher frequency there is no correlation. We have therefore averaged all $g_{\alpha{\rm PE}}(r)$ in the rage $0 \leq \omega \leq \omega_{\rm max}$ in order to get a correlation function that takes into account all the TDs at low frequencies. Since the number of TDs increases quadratically with $\omega$, see Fig.~\ref{fig2_struTD}{\bf (d)}, we have weighted the individual $g_{\alpha{\rm PE}}$ by $\omega^{-2}$. The resulting correlation functions $g_{+{\rm PE}}^{\rm av}(r)$ and $g_{-{\rm PE}}^{\rm av}(r)$ are shown in panel {\bf (c)} and from this graph one clearly recognizes that the negative TDs are significantly correlated with the PE in that the peak at small $r$ rises above 2.5, which means that there is a more than two-fold increase of the probability that a PE occurs close to a TD as compared to a uniform distribution.

Since these results have been obtained by considering all the modes up to a cut-off frequency $\omega_{\rm max}=3.5$, it is of interest to check to what extent the correlation between the TDs and PE depend on the frequency of the mode. Therefore we show in Fig.~\ref{fig3_TDPE}{\bf (d)} the probability that a PE is close to a TD as a function of $\omega$. In practice we used a cut-off distance between TD and PE of 1.6, i.e., the location of the first minimum in the radial distribution function between the particles, see Extended Data Fig.~\ref{fig_s9_g_of_r}. The graph shows that the probability is high for small value of $\omega$ and then decreases rapidly if one approaches the threshold 3.5, the value we have used to calculate the curves in panel {\bf (c)}, thus justifying this choice. For larger values of $\omega$ the probability stays at around 0.3, a value that is determined by the spatial density of the PE, i.e., these high frequency modes are not relevant for the occurrence of the PEs.

The results presented so far are for a strain of $\gamma=0.025$. Also included in panel~{\bf (d)} is the corresponding data for $\gamma=0.05$, a value that is close to the yielding point of the system, see Extended Data Fig.~\ref{fig_s4_stress_strain}, and one recognizes that also in this case there is a pronounced increase of the probability at small $\omega$, i.e., the negative TDs for small $\omega$ correlate well with the PE and this correlation disappears for $\omega \gtrsim 3.5$, indicating that this value is independent of the strain. Note that this threshold is also close to the frequency at which the number of TDs as a function of frequency starts to show the first deviation from the quadratic law found at low values of $\omega$, see Fig.~\ref{fig1_TD}{\bf (f)}. Furthermore it also coincides with the frequency at which the density of states starts to show deviations from the Debye regime, see Extended Data Fig.~\ref{fig_s2_vdos}{\bf (b)}, i.e., above this $\omega$ the nature of the vibrational modes is no longer acoustic and starts to become affected by the disorder of the structure. Therefore one concludes that this threshold does indeed reflect a relevant frequency for the plastic yielding of the sample. 

In order to have a visual impression of the correlation between the location of the TDs and the PEs we present in Fig.~\ref{fig3_TDPE}({\bf e)} the charge density field of these TDs (color map) as well as the location of the PEs (symbols). The graph demonstrates the strong correlation between zones with a high density of $-1$~TDs and the PEs. (This strong correlation becomes even more visible if we just consider the $-1$ TDs to generate the colormap, see Extended Data Fig.~\ref{fig_s10_num_dens_field}.) One also notices that basically all zones of high density do contain PEs, i.e., the negative TDs are indeed able to predict the location of the ``soft spots'', the points at which the sample is yielding. Note that we have done the shearing in the $+x$ and the $-x$ direction and, as expected, the location of the corresponding plastic events (filled and open symbols) are not quite the same. However, both types of PEs do show a good correlation with the zones of strong negative charge, i.e., to a first approximation the tensorial character of the plastic response can be neglected. One also recognizes clearly the anti-correlation between the PE and the positive TDs (zones in red), showing that the former are forming ``hard spots'' in the samples that are stable under a shear transformation. This result is reasonable since the positive TDs correspond to the points of the vibrational excitations at which the vibrational amplitude goes smoothly to zero and hence it can be expected that the nearby particles will not be affected strongly by the external shear. This rational is consistent with earlier findings that particles with larger amplitude of the vibrational eigen-vector contribute more to the non-affine deformation and thus irreversible rearrangements~\cite{Din14,Man11}.

{\it Discussion:}
The identified correlations between the location of the topological defects and the plastic events indicate that these quantities are intimately related to each other. Zones of the sample which have a high density of defects with topological charge $+1$ are stable towards shearing since in their vicinity the amplitude of the vibrational excitations are small or vary smoothly in space, i.e., the application of a local strain will not affect significantly these vibrational modes and thus will not give rise to a plastic event. This is also the case for the boundary between two neighboring regions of +1 TD with opposite chirality, since also then the flow field is a smooth function in space, see Fig.~\ref{fig1_TD}{\bf (e)}. For zones with a high density of TD with topological charge $-1$, the vector field of the eigen-modes will instead be strongly disturbed by the presence of a local strain, thus making it is highly probable for a plastic event to occur. Note that this connection between TD and PE can be expected to hold only if the local shear is indeed a smooth function in space, i.e., if the sample can locally be described as an elastic medium. This is only the case if the considered length scales are sufficiently large, which in turn rationalizes our finding that the correlation between PE and TD is pronounced if the distance between the TDs is larger than the lengths scale on which the system can support acoustic phonons, while for smaller scales the correlation is lost. Our results also shed light on the previous studies in which machine learning approaches have been used to predict the location of PE~\cite{Bap20,Fon22,Sch17,Cub17,Fan21,Rid22}. Despite the success of these approaches, the understanding on the physical origin of these soft spots has so far not really been elucidated and our findings do now allow us to interpret these zones in terms of a simple physical quantity. Note that in contrast to previous studies, Refs.~\cite{Mal04,Mal04a,Bon19,Bag21}, we do not have to determine the normal modes as a function of the strain in order to predict the location of the PE. Instead this location is obtained directly from the normal modes at strain zero, i.e., it is a purely structural quantity in the unperturbed system. This feature should thus  permit us to make headway with analytical calculations that aim to describe the yielding of amorphous systems and hence to obtain a better understanding of this important problem. Finally we mention that the approached presented here can be generalized in a straightforward manner to three dimensional systems and it will thus be highly interesting to determine for these cases the correlation between PE and TD.

\newpage

{\bf Methods}\\
\noindent{\bf MD simulation.} The two-dimensional glass former we study 
is an equimolar binary mixture of particles with size $\sigma_1$ and $\sigma_2$ that interact via a truncated Lennard-Jones potential:

\begin{equation}
u_{ab}(r)=4\varepsilon\left[\left(\frac{\sigma_{ab}}{r}\right)^{12}-\left(\frac{\sigma_{ab}}{r}\right)^6\right]+C_{ab} \quad .
\label{mdpotential}
\end{equation}

\noindent
Here $\sigma_{12}=(\sigma_1+\sigma_2)/2$ and the constant $C_{ab}$ ensures that $u_{ab}$=0 at $r_{cut}=2^{1/6}\sigma_{ab}$, so the potential is purely repulsive and continuous at the cutoff distance. The size ratio was set to 1.414 to prevent crystallization~\cite{Ham06,Tan10}. A total of $N$=10000 (or $N$=1250) particles with the mass ratio  $m_1/m_2=(\sigma_1/\sigma_2)^2$ were enclosed in a square box of length $L$=115.47 (number density is 0.75) with periodic boundary conditions. The units of length, mass, and energy are $\sigma_1$, $m_1$, and $\varepsilon$, respectively. Time and temperature are in units of $\tau=\sigma_1\sqrt{m_1/\varepsilon}$ and $\varepsilon/k_B$, with $k_B$ the Boltzmann constant. The time step of integration was 0.001$\tau$. We first equilibrated the system in the liquid state at $T$=5.0 during $10^6$ MD steps and then quenched it to the final temperature $T$=0.1 within $10^7$ time steps using a linear cooling schedule, after which we annealed the sample for another $10^6$ MD steps. ($T_{\rm MCT}$=1.09 for our system, see Extended Data Fig.~\ref{fig_S5_diffusion}.)\\

\noindent{\bf Normal modes.} The zero-temperature glass was generated by a  conjugate gradient energy minimization process, and the vibrational normal modes were obtained by diagonalizing the full dynamical matrix.\\

\noindent{\bf Topological defects.} To identify the topological defects of an eigenvector field $(e_i^x,e_i^y)$, $i=1,\ldots 2N$, we first assigned an angle $\theta(\vec{r})$ on every site $\vec{r}$ of a 100$\times$100 square lattice superposed to the sample. (We have checked that a 120$\times$120 grid does not change the main conclusions presented in the main text.) This angle $\theta(\vec{r})$ is defined to be

\begin{equation}
\tan\theta(\vec{r})=\sum_i w(\vec{r}-\vec{r_i})e_i^y/\sum_i w(\vec{r}-\vec{r_i})e_i^x \quad ,
\label{tantheta}
\end{equation}

\noindent
where $\vec{r_i}$ is the location of particle $i$, and $w(\vec{r}-\vec{r_i})$ is a weight function (in practice a Gaussian~\cite{Oya19,Gol05}: $w(\vec{r}-\vec{r_i})=\exp(-|\vec{r}-\vec{r_i}|^2/\sigma_1^2)$). Hence this procedure allows to define a map from the eigenvector field, given at the positions of the particles, to a regular grid. The topological defects were then identified by integrating $\theta$ over a closed path on the lattice, and located by taking the position of the center of the smallest square with a value of $2\pi$ ($-2\pi$) of this integral, defining thus charges of +1 and $-1$ respectively.\\

\noindent{\bf Plastic events.} The athermal and quasistatic shear was realized by shearing the simulation box by a small strain increment $\Delta\gamma$ 
and subsequently minimizing the energy of the configuration with a conjugate gradient algorithm. This procedure was repeated until the global strain reached $\gamma =2.5$\%, using $\Delta \gamma=0.05$\%, (and 5\%, using $\Delta \gamma=0.1$\%), the strain at which the data in Fig.~\ref{fig3_TDPE} is shown.
The resulting stress-strain curve is presented in Extended Data Fig.~\ref{fig_s4_stress_strain} for the strain in the $+x$ and the $-x$ direction. This figure also demonstrates that if the strain exceeds 7-8\%, the average strain does not grow anymore, i.e., that the sample has yielded.

The locally irreversible rearrangements of particles in
the glass are evaluated by the non-affine displacement 
$D_{\rm min}^2$ introduced in Ref.~\cite{Fal98} and which is defined as the minimum of

\begin{equation}
D_i^2=\frac{1}{N_i}\sum_{j(i)}\left[\vec{r}_j(\gamma)-\vec{r}_i(\gamma)-{\bf F}_i\times
(\vec{r}_j(\gamma-\Delta\gamma)-\vec{r}_i(\gamma-\Delta\gamma))\right]^2,
\label{D2}
\end{equation}

\noindent
where the index $j(i)$ runs over all the particles that are nearest neighbors of particle $i$ (thus the distance is less than 1.6, the first minimum
of the total pair correlation function), $\vec{r_j}(\gamma)$ denotes the position of particle $j$ after the minimization of the box that is sheared by $\gamma$, and $N_i$ is the number of nearest neighbors of particle $i$. The matrix ${\bf F}_i$ is chosen such that it minimizes $D_i^2$ and it can be calculated based on the spatial coordinates of the corresponding particles~\cite{Fal98}.\\

{\noindent{\bf Correlation function between the TDs and and the PEs.} 
For each eigenmode $\kappa =1,2,..., 2N$, we define the radial pair correlation function $g_{\kappa,\alpha {\rm PE}}(r)$ between the TDs and PEs, with $\alpha \in \{-1,+1\}$, as

\begin{equation}
g_{\kappa,\alpha \rm PE}(r)=\frac{L^2}{2\pi r N_{\rm TD}N_{\rm PE}}
\sum_{i=1}^{N_{\rm TD}}\sum_{j=1}^{N_{\rm PE}}\delta(r-|\vec{r}_{ij}|) \quad.
\label{grtdpe}
\end{equation}

\noindent
Here $N_{\rm TD}$ and $N_{\rm PE}$ are the number of TD of the mode $\kappa$ and the number of PEs, respectively, and $r_{ij}$ is the distance between the TD $i$ and the PE $j$.

The average correlation function $g_{\alpha \rm PE}(r)$ is then given by

\begin{equation}
g_{\alpha \rm PE}(r)=\frac{\sum_\kappa g_{\kappa,\alpha\rm PE}(r)/\omega_\kappa^2}{\sum_\kappa 1/\omega_\kappa^2} \quad ,
\label{grtdpe2}
\end{equation}

\noindent
where the sum over $\kappa$ runs up to the cut-off frequency $\omega_{\rm max}$ defined in the main text.
}
\vspace*{10mm}

{\bf Acknowledgments}\\
We thank M.Z. Li, P.F. Guan, and the other members of the “Beijing Metallic Glass Club”
for the long-term fruitful discussions. W.K.~is senior member of the Institut universitaire de France. This work was supported by the National Natural
Science Foundation of China (Grant Nos. 11804027, 11935002, and 52031016).\\

{\bf Author contributions:} ZWW, WK, and LX designed the project with advice from WHW. ZWW carried out the simulations, and YC assisted in the identification and analysis of the TDs. ZWW, WK, LX, and WHW analyzed the data. ZWW, WK, and LX wrote the paper.\\

\bibliographystyle{naturemag}

\newpage

\renewcommand{\figurename}{{\bf Extended Data Fig.}}
\setcounter{figure}{0}
\setstretch{1.2}
\nolinenumbers

\begin{figure}[ht]
\centering
\includegraphics[width=0.8\linewidth]{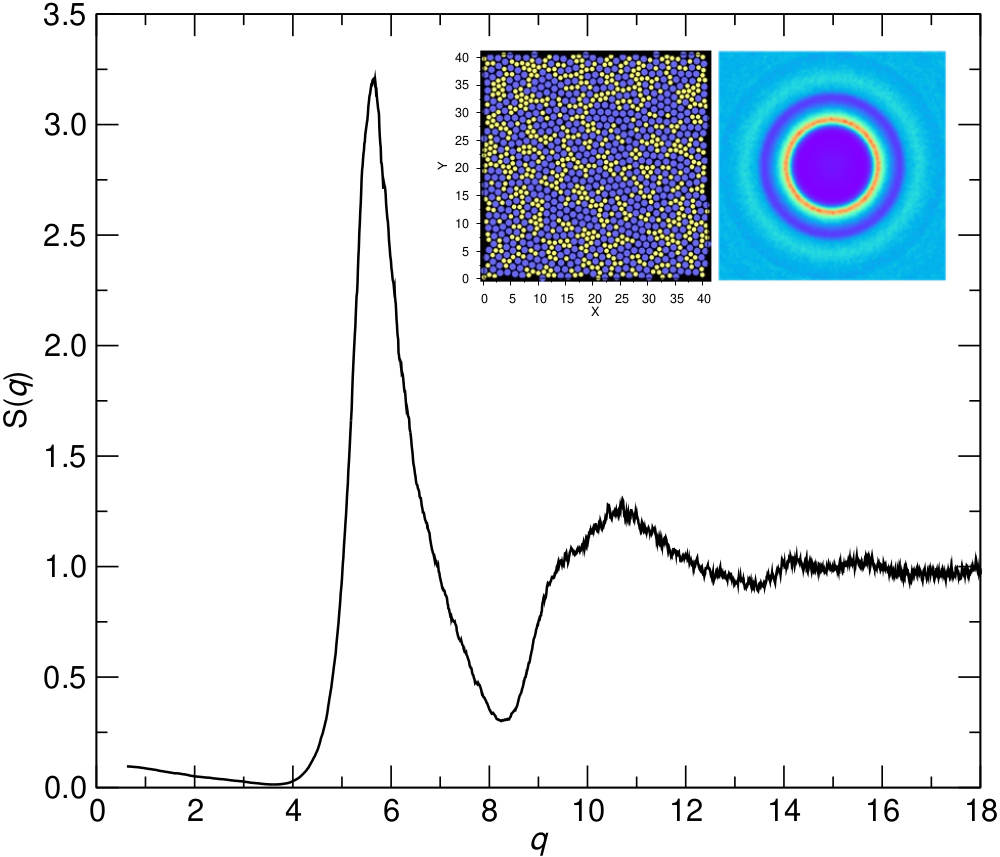}
\caption{
{\bf The static structure factor $S(q)$ of the model glass ($N$=1250) at a number density of 0.75 at $T$=0.1.} $S(q)$,
calculated from the position of the particles via $S(q)=(1/N)\langle |\sum_je^{-{\rm i}{\bf q}\cdot{\bf r}_j} |^2\rangle$, shows the typical form found in disordered systems. The left inset shows a snapshot of the system and the right inset shows the two-dimensional static structure factor.
\label{fig_s1_s_of_q}
}
\end{figure}

\begin{figure}[ht]
\centering
\includegraphics[width=0.8\linewidth]{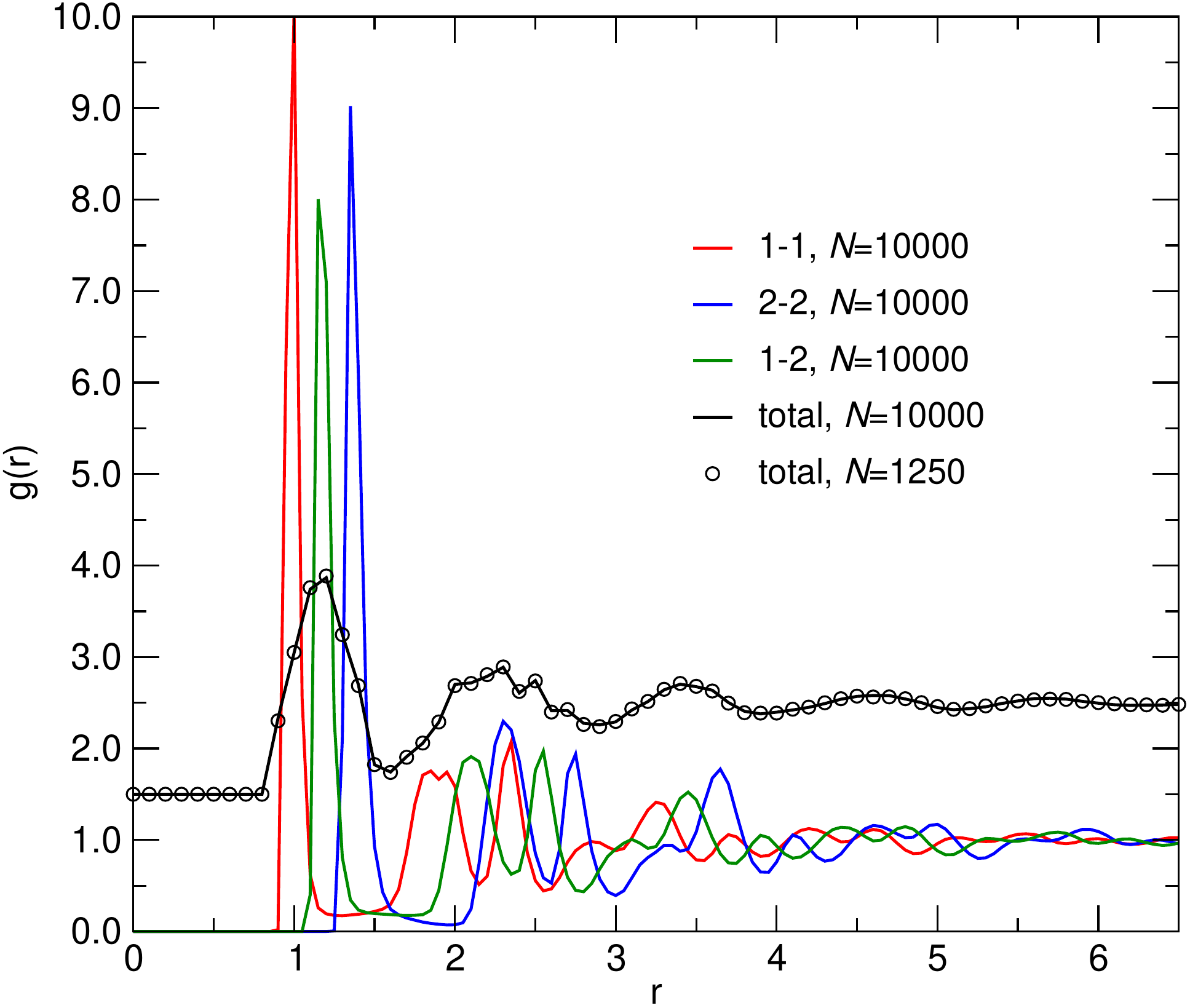}
\caption{
{\bf Radial distribution function.} The colored curves are the partial radial distribution functions as obtained for the system with $N=10^4$ particles. The resulting total radial distribution function is show in black (shifted vertically by 1.5 for the sake of visibility). The back symbols are the total radial distribution function for the system with $N=1250$ particles. It superposes perfectly with the data for the larger system, showing that there are no finite size effects. 
}
\label{fig_s9_g_of_r}
\end{figure}

\begin{figure}[ht]
\centering
\includegraphics[width=0.8\linewidth]{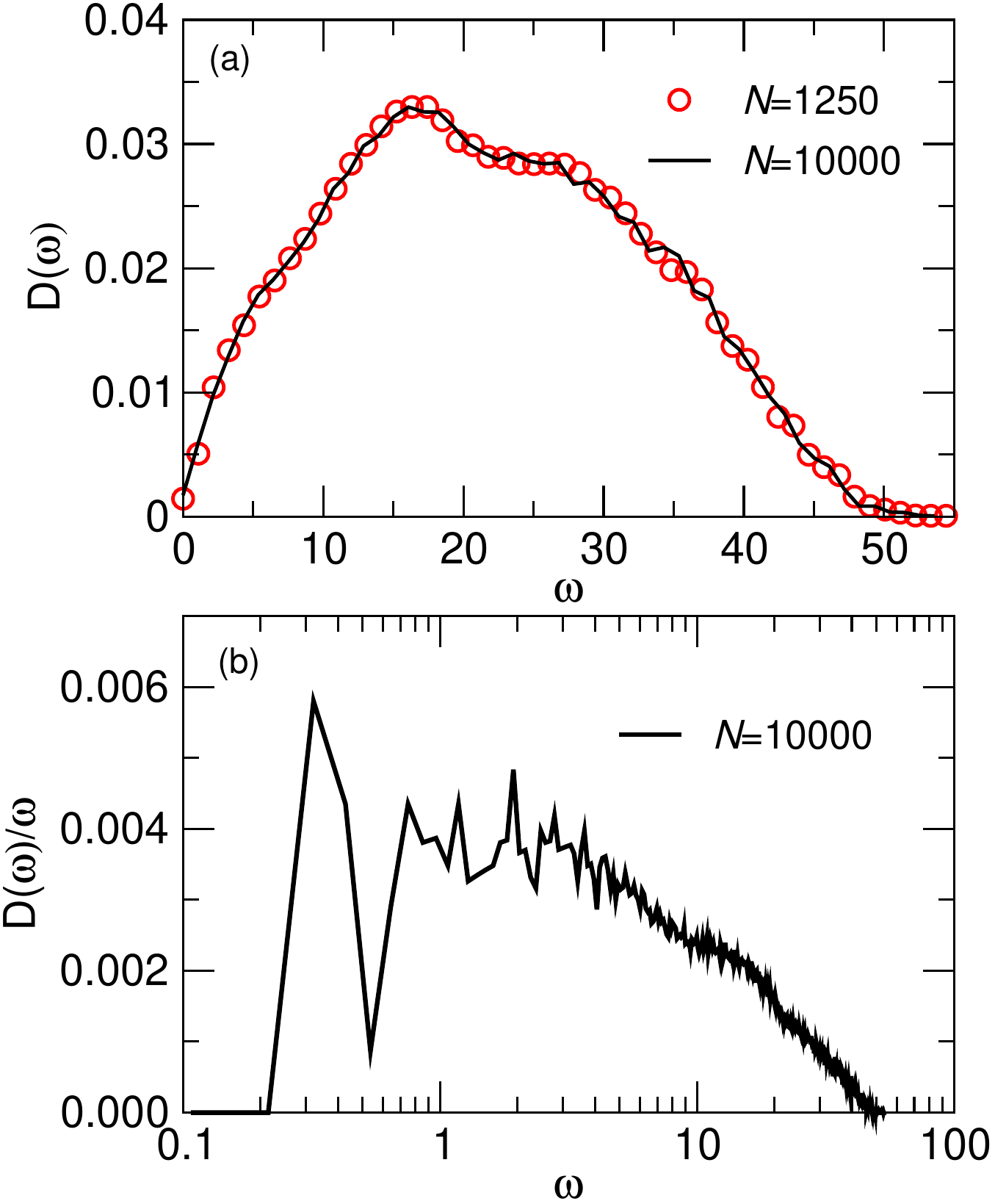}
\caption{
{\bf Vibrational density of states.}
(a): Vibrational density of states for two system sizes. No system size dependence is seen. For $N$=1250, the data has been averaged over 12 independent samples.
(b): Vibrational density of states divided by the frequency. At low frequencies this ratio is a constant, thus showing that the Debye law holds at low $\omega$ while it breaks down for frequencies above $\omega \approx 3.0$.
\label{fig_s2_vdos}
}
\end{figure}

\begin{figure}[ht]
\centering
\includegraphics[width=0.8\linewidth]{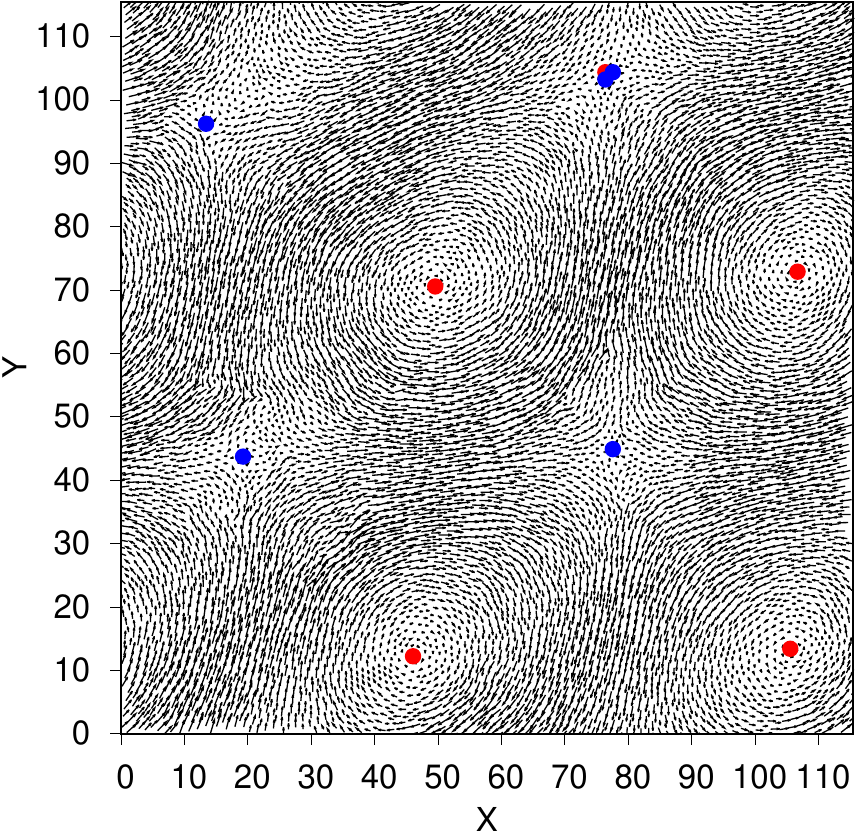}
\caption{
{\bf Ordering of the eigen-vector field and topological defects at low frequencies.}  $N$=10000 and $\omega$=0.51.
At small frequencies the vibrational modes are of acoustic nature and hence form a regular pattern, the geometry of which depends on the eigen-mode considered. As a consequence the resulting topological defects show also an ordered arrangement (+1~TDs in red and $-1$~TDs in blue).
}
\label{fig_s3_ev}
\end{figure}

\begin{figure}[ht]
\centering
\includegraphics[width=0.8\linewidth]{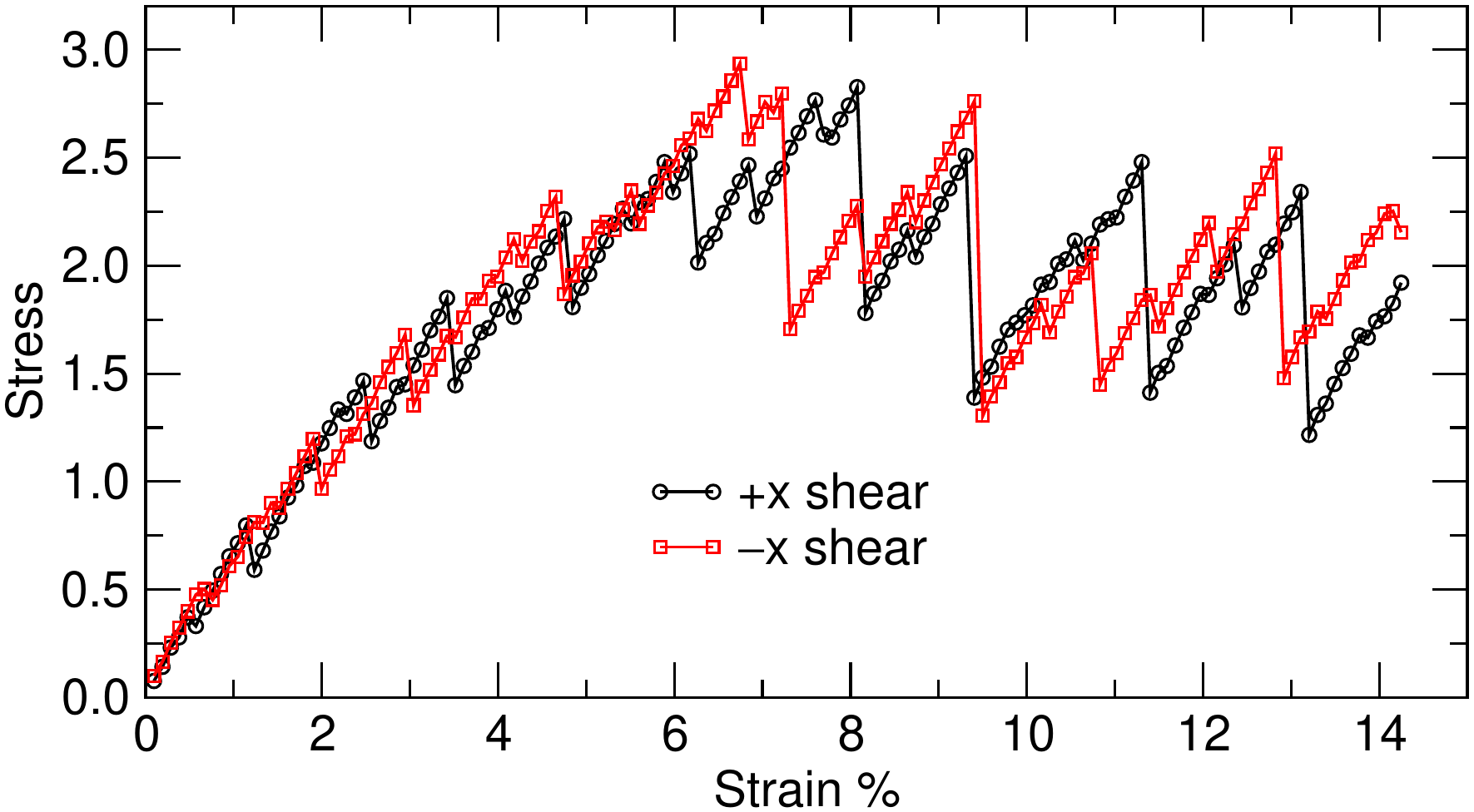}
\caption{
{\bf Athermal quasi-static strain-stress curves for the system with $N$=10000.}
In this system, up to a global strain of 2.5\%, we have several significant drops in the strain-stress curve, which gives us the plastic events shown in Fig.~\ref{fig3_TDPE}{\bf (e)} of the main text. For strains larger than $\approx 7$\% the stress fluctuates around a mean value, indicating that the sample has yielded. Note that the two shear directions give statistically the same behavior.
}
\label{fig_s4_stress_strain}
\end{figure}

\begin{figure}[ht]
\centering
\includegraphics[width=0.65\linewidth]{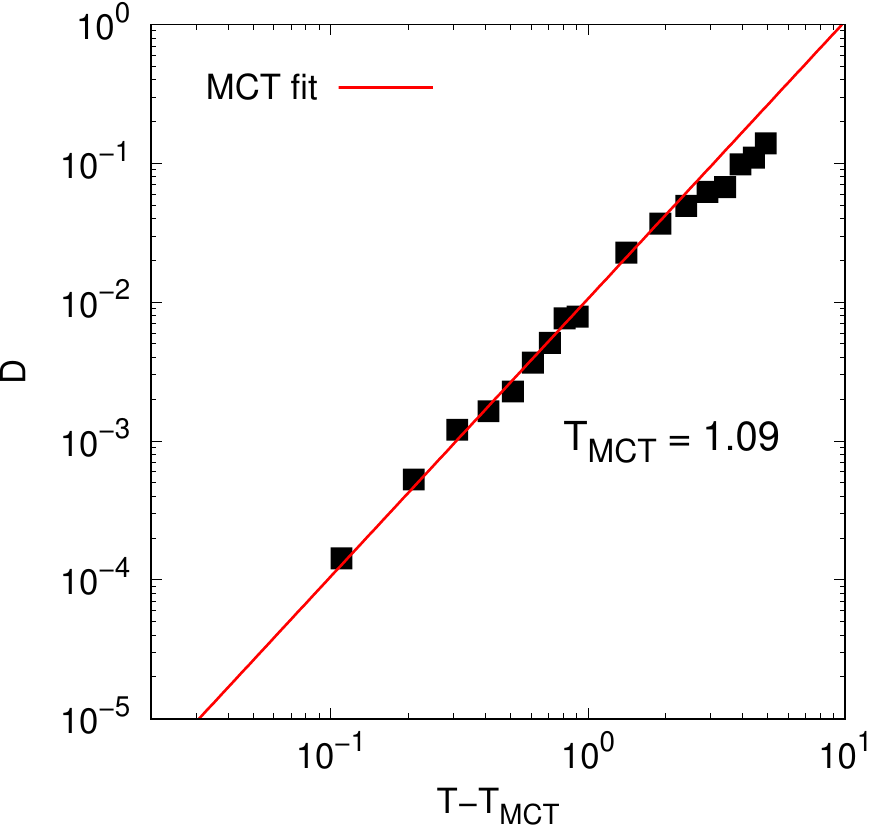}
\caption{
{\bf Temperature dependence of the diffusion constant $D$.} 
The diffusion constant $D$ has been obtained from the mean squared displacement of the particles and the Einstein relation. To determine $D$ we quenched the system to the target temperature, equilibrated it for $10^6$ steps and then made a production run of length $2\times 10^6$ steps. Also included in the graph is a fit with a power-law suggested by mode-coupling theory
(MCT) of the form $D \propto (T-T_{\rm MCT})^\gamma$, where $T_{\rm MCT}$ is the critical temperature of the theory. The so obtained values for $T_{\rm MCT}$ and $\gamma$ are $T_{\rm MCT}=1.09$ and $\gamma=2.0$, respectively. Thus the temperature $T=0.1$ at which we probe our glass sample is deep in the glass state.
}
\label{fig_S5_diffusion}
\end{figure}

\begin{figure}[ht]
\centering
\includegraphics[width=0.6\linewidth]{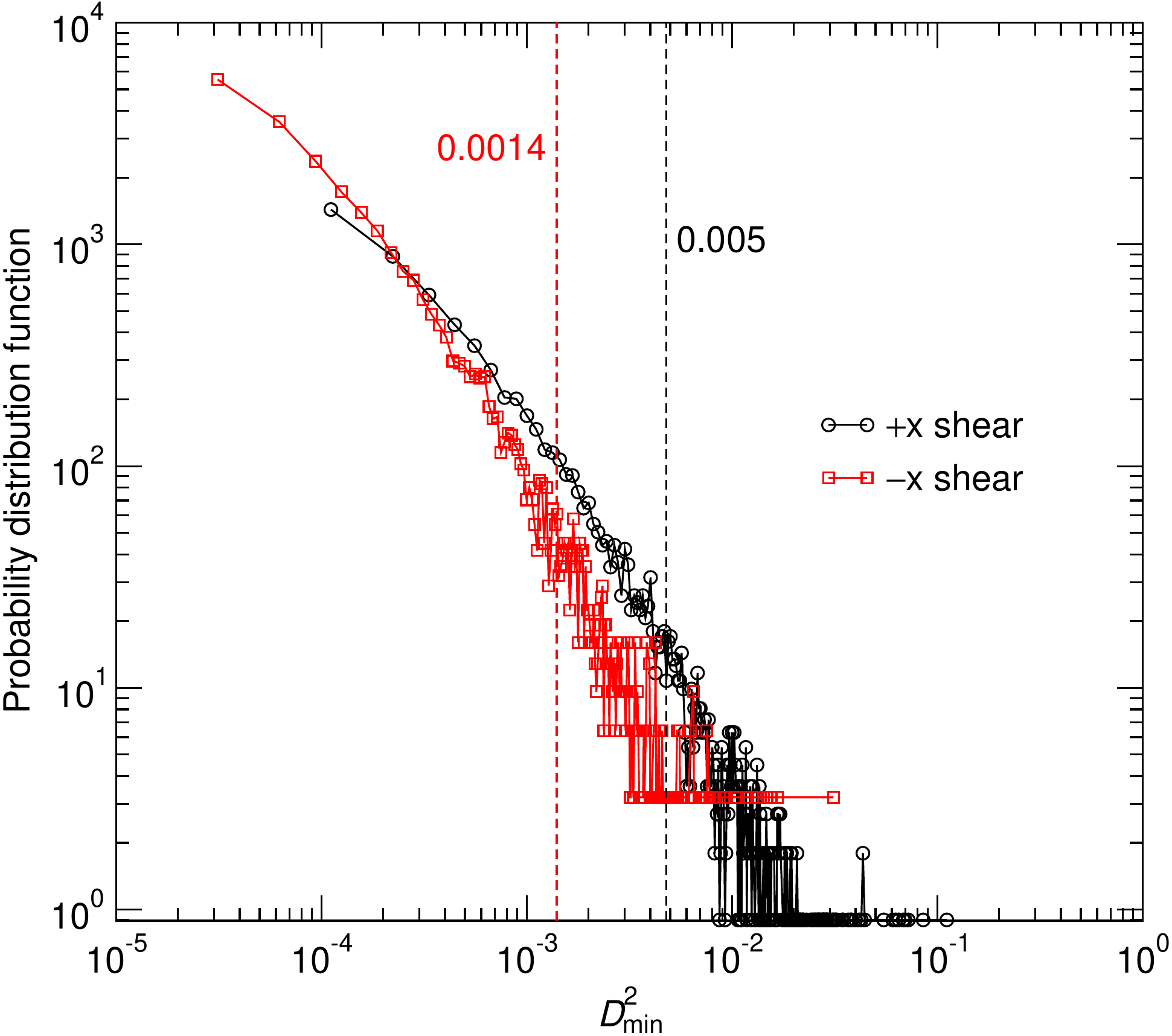}
\caption{
{\bf Probability distribution function of $D^2_{\rm min}$.}
Distribution of the non-affine displacement $D^2_{\rm min}$ after the system has been sheared to the strain 2.5\%. The two curves correspond to a shear in the $+x$ and $-x$ directions. Particles that are in the top 5\% of the displacement are used to define the plastic events shown in Fig.~\ref{fig3_TDPE} of the main text. This corresponds to values of $D^2_{\rm min}> 0.005$ for the $+x$ shear and $0.0014$ for the $-$x shear, respectively, see vertical dashed lines. The difference of these two values are due to finite size effects.
}
\label{PDF_d2}
\end{figure}

\begin{figure}[ht]
\centering
\includegraphics[width=1.0\linewidth]{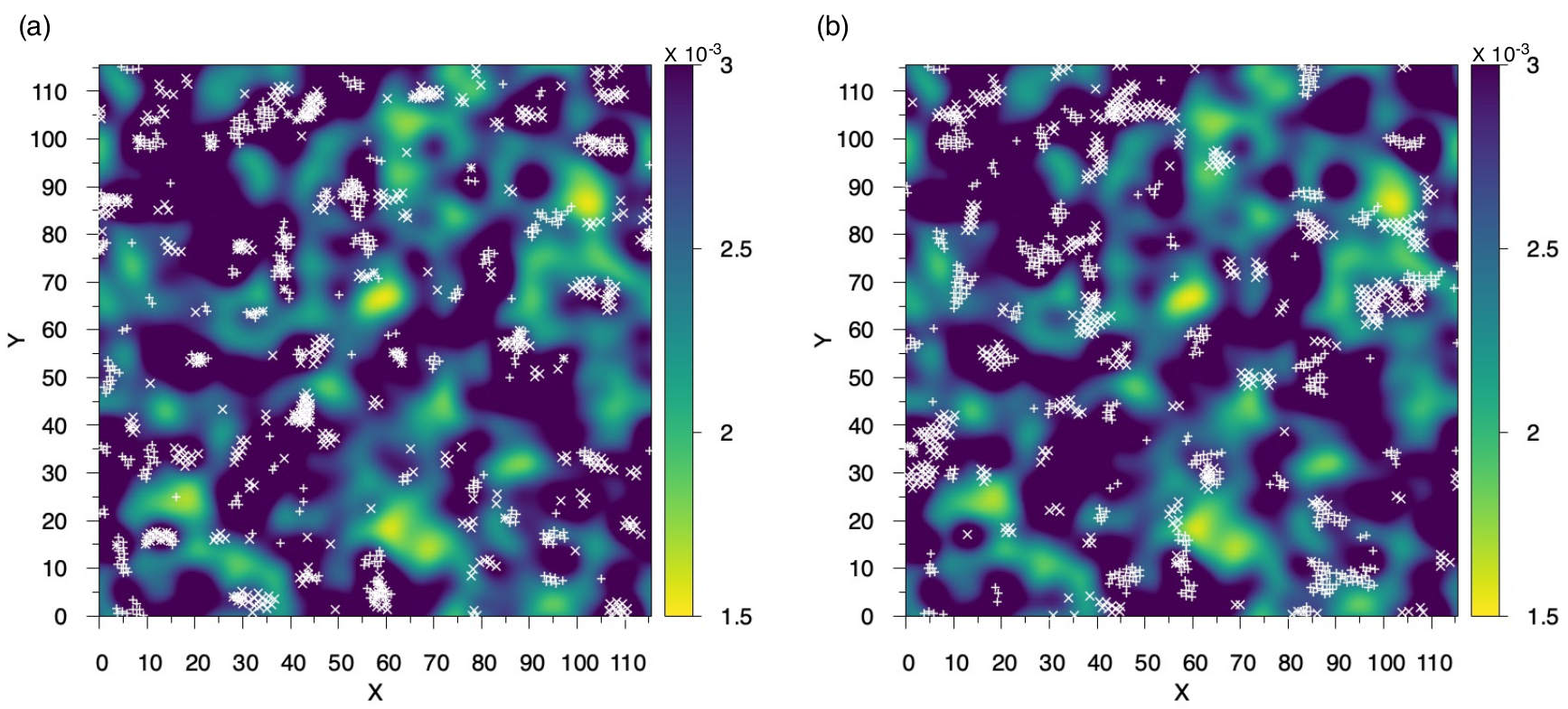}
\caption{
Number density field of the $-1$ TDs (color map) and of the PE (symbols) at strain $\gamma=2.5\%$, (a), and $\gamma=5\%$, (b). The density field, in units of TD per unit area, has been obtained by averaging the TDs of all the modes over all frequencies up to $\omega_{\rm max}=3.5$, using a weight factor $\omega^{-2}$, and a subsequent Gaussian smoothing function of width 4. PE resulting from a shear in the positive and negative $x$-direction are marked by $+$ and $\times$ filled symbols, respectively. The comparison of the two panels shows that an increase of strain does not modify significantly the pattern of the PE.}
\label{fig_s10_num_dens_field}
\end{figure}

\end{document}